\documentclass{ckm}                 

\usepackage{txfonts}            

\newcommand{\dmd}{\Delta m_d}
\newcommand{\etal}{\begingroup\it et al.\endgroup}

\confname{Workshop on the CKM Unitarity Triangle, IPPP Durham, April 2003}

\title{Advances in $\dmd$ measurements}

\author{F J Ronga}
\address{IPHE, University of Lausanne}

\begin{document}

\begin{abstract}
We report the current status of $\dmd$ measurements at $B$-factories. The most
recent world average is $\dmd = 0.502\pm0.007$~ps${}^{-1}$ (1.4\%
accuracy). An estimate of the errors for 500~fb${}^{-1}$ data is also given.
\end{abstract}

\maketitle

\section{Introduction}

In the Standard Model, $B^0$-$\overline B{}^0$ oscillations occur through
second-order weak interactions, mainly through internal loops containing
virtual $t$ quarks. The mixing parameter $\dmd$, the mass difference between
the two mass eigenstates, is thus related to the $V_{tb}$ and 
$V_{td}$ CKM matrix elements. The measurement of $\dmd$ can therefore in 
principle provide a means to extract $|V_{td}|$. In addition, $\dmd$ plays a 
role in the parameterization of the $CP$ asymmetries in the $B^0$ system: 
a precise measurement of $\dmd$ is also needed for $CP$ violation 
measurements (see~\cite{bab_sin2phi1,bel_sin2phi1}).

In this article, we present different measurements of $\dmd$ from the 
time distributions of 
opposite-flavor (OF -- $B^0\overline B{}^0$) and same-flavor 
(SF -- $B^0B^0$, $\overline B{}^0\overline B{}^0$) neutral $B$ decays 
at the $\Upsilon(4S)$ resonance. The theoretical time-dependent 
probabilities for observing OF and SF states are given by:
\begin{eqnarray}\label{prob_func}
{\cal P}^{\rm OF}(\Delta t) &=& \frac{e^{-|\Delta t|/\tau_{B^0}}}{4\tau_{B^0}}
     [1+\cos(\dmd\,\Delta t)]\nonumber\\
{\cal P}^{\rm SF}(\Delta t) &=& \frac{e^{-|\Delta t|/\tau_{B^0}}}{4\tau_{B^0}}
     [1-\cos(\dmd\,\Delta t)]\nonumber\\
\end{eqnarray}
where $\tau_{B^0}$ is the $B^0$ lifetime and $\Delta t$ is the proper
time difference between the two $B$ meson decays. This assumes CP and CPT
conservation in the mixing, as well as negligible $\Delta\Gamma$
(decay width difference between the two $B$ meson mass eigenstates).

The analyses presented here were performed on data collected with the
BaBar  and the Belle detectors~\cite{detectors}. A 9~GeV (resp. 8~GeV)
electron beam  and a 3.1~GeV (resp. 3.5~GeV) positron beam are
collided in the PEPII (resp. KEKB) storage ring, resulting in a
Lorentz boost of the center-of-mass  of $\beta\gamma=0.55$
(resp. $0.425$) with respect to the laboratory frame. Since $B$
mesons are nearly at rest in the $\Upsilon(4S)$~frame, the proper time
difference $\Delta t$ is approximated by $\Delta z/\beta\gamma c$,
$\Delta z$ being the (signed) distance between the decay vertices of
the two $B$ mesons along the beam axis. The flavor of the $B$ mesons
is determined using flavor-specific decays.

\section{Dilepton measurement}

In this analysis~\cite{bab_dilep,bel_dilep}, semi-leptonic decays
$B\rightarrow X^-\,l^+\,\nu_l$ from both $B$ mesons are used to tag
the $B$ flavor with the sign of the  lepton. Two fast leptons are
searched for.  The decay vertex position of the $B$ meson is
determined from the interception of the lepton track with the profile
of the interaction point. Because of the large semi-leptonic branching
fraction, this analysis offers the largest statistics. However, the
purity of the signal is affected by the background coming from charged
$B$ mesons.

\begin{figure}[htb]
\includegraphics*[width=\hsize]{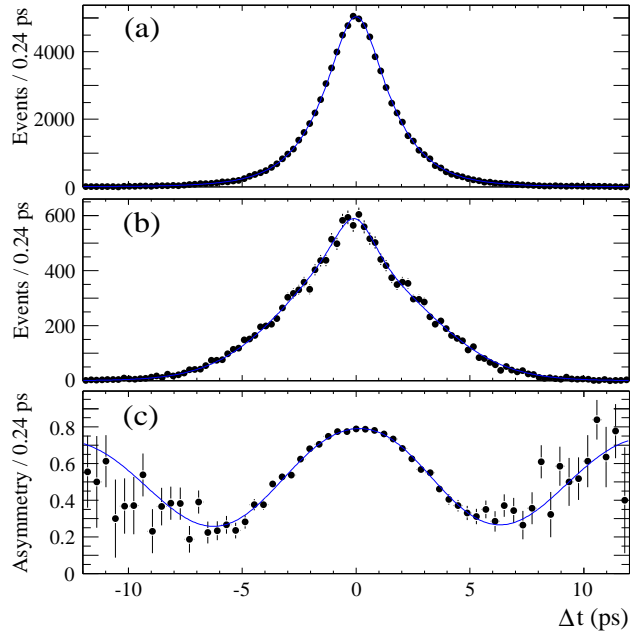}
\caption{Result of the fit for the dilepton measurement by BaBar on
(a)~OF and (b)~SF events, and the corresponding mixing
asymmetry $(N_{\rm OF} - N_{\rm SF})/(N_{\rm OF} +
N_{\rm SF})$.}
\label{di_babar}
\end{figure}

\begin{figure}[htb]
\includegraphics*[width=0.8\hsize]{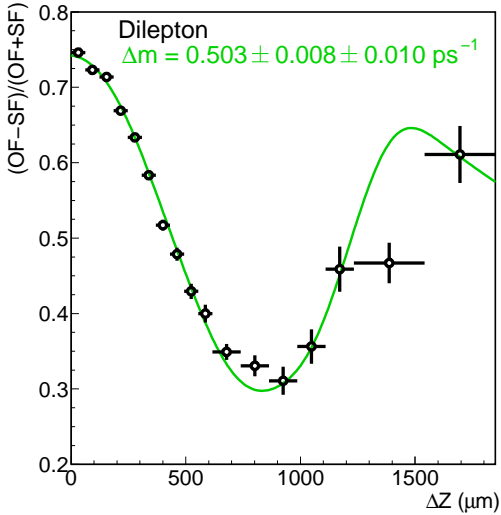}
\caption{Mixing asymmetry fitted by Belle on the dilepton $\Delta z$ 
distributions.}
\label{di_belle}
\end{figure}

This selection has been applied on 20~fb${}^{-1}$ on-resonance data
from BaBar (about 22 million $B$ meson pairs), yielding $\dmd =
0.493\pm0.012\pm0.009$~ps${}^{-1}$ (see Fig.~\ref{di_babar}, also
showing the mixing asymmetry  $(N_{\rm OF} - N_{\rm SF})/(N_{\rm OF} +
N_{\rm SF})$). Belle obtains $\dmd = 0.503\pm0.008\pm0.010$~ps${}^{-1}$
from 29.4~fb${}^{-1}$ on-resonance data~(see Fig.~\ref{di_belle}). At
present, the latter  is the most precise single measurement of $\dmd$
(with an accuracy of about 2.5\%). 

\section{\boldmath$ D^*\,\pi$ partial reconstruction}

Belle uses another partial reconstruction method~\cite{yangheng} to extract
$\dmd$ from $B^0\rightarrow D^{*-}\pi_f^+$ decays. The $D^{*-}$
information is extrapolated from the soft pion of $D^{*-}\rightarrow
\overline D{}^0\pi^-_s$, and then combined with the fast pion
$\pi_f^+$ and the beam information to reconstruct the $B$ meson. The
flavor is given by the charge of the fast pion. The other side is
tagged by simply  looking for a fast lepton. 31~fb${}^{-1}$ 
on-resonance data was used in this measurement.

\begin{figure}[htb]
\includegraphics*[width=\hsize]{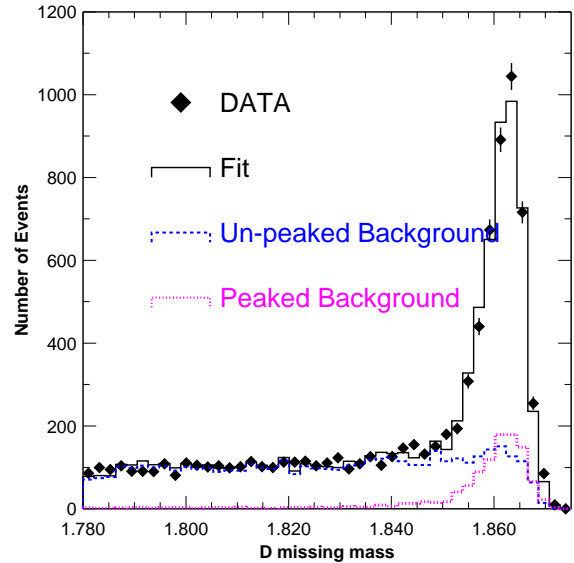}
\caption{The $D^0$ ``missing mass'' fitted with Monte-Carlo data for
the $D^*\,\pi$ measurement (Belle).}
\label{d0mms}
\end{figure}

Contributions of various backgrounds can be estimated from the $D^0$
``missing mass'' (see Fig.~\ref{d0mms}). A simultaneous unbinned
maximum likelihood fit to OF and SF events yields:
$\dmd=0.509\pm0.017\pm0.020$~ps${}^{-1}$.

\section{\boldmath $D^*\,\ell\,\nu$ full reconstruction}

This method~\cite{bab_dslnu,bel_dslnu} fully reconstructs $B^0\rightarrow D^{*-}\,\ell^+\,\nu$, with 
$D^{*-}\rightarrow \overline D{}^0\pi^-$ and $\overline D{}^0\rightarrow K^+\pi^-$,
$K^+\pi^-\pi^0$, or $K^+\pi^-\pi^+\pi^-$ (BaBar also reconstructs $K_s\pi^+\pi^-$). The
flavor of the other $B$ meson is tagged using the same algorithms as
in Ref.~\cite{bab_sin2phi1,bel_sin2phi1} (a neural network for BaBar, a
multidimensional likelihood for Belle).

\begin{figure}[htb]
\includegraphics*[width=\hsize]{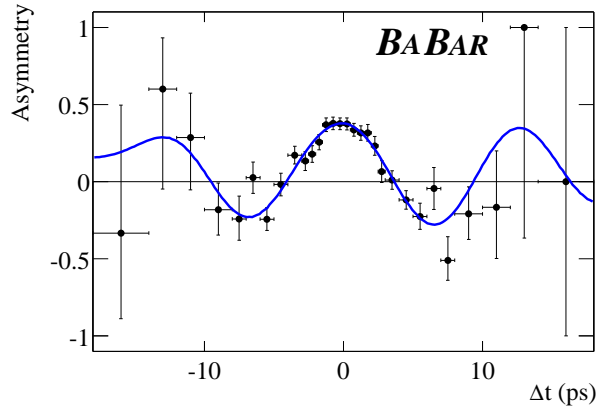}
\caption{Mixing asymmetry showing the data points and the overlaid
fit to $\Delta z$ distributions from the $D^*\,\ell\,\nu$ analysis (BaBar).}
\label{dslnu_fit}
\end{figure}

The cosine of the angle between the reconstructed $D^*\,\ell$ system and the 
$B$ meson momenta in the $\Upsilon(4S)$ frame is used to separate the
signal from various backgrounds ($D^{**}$, fake $D^*$, random
$D^*\,\ell$ and continuum).
An unbinned maximum likelihood fit is then performed on the $\Delta z$
distributions. The results of the fit by BaBar, shown on
Fig.~\ref{dslnu_fit}, yields $\dmd = 0.492\pm0.018\pm0.013$~ps${}^{-1}$ 
(from 20~fb${}^{-1}$). Belle obtains $\dmd = 0.494\pm0.012\pm0.015$~ps${}^{-1}$ 
from 29~fb${}^{-1}$.

\section{Hadronic modes}

The exclusive reconstruction of $B$ mesons decaying into flavor
specific hadronic states has also been used to measure
$\dmd$~\cite{bab_hadron,bel_hadron}. Neutral $B$ mesons are reconstructed in the
decay modes $D^{(*)+}\pi^-$ and $D^{(*)+}\rho^-$ (and also $D^{(*)+}a_1^-$
by BaBar), with $D^{*+}$ decaying into $D^0\pi^+$ and $D^0$ decaying into
$K^-\pi^+$, $K^-\pi^+\pi^0$ or $K^-\pi^+\pi^-\pi^+$ (or also
$K_s\pi^+$ in BaBar's case). The $\rho^-$ and $a_1^-$ are formed of
$\pi^0\pi^-$ and $\pi^-\pi^+\pi^-$, respectively. Finally, BaBar also
includes $B^0$ decaying into $J/\psi K^{*0}$. The other side is tagged
using a flavour tagging algorithm, as in the previous analysis
(see~\cite{bab_sin2phi1,bel_sin2phi1}).

\begin{figure}[htb]
\includegraphics*[width=\hsize]{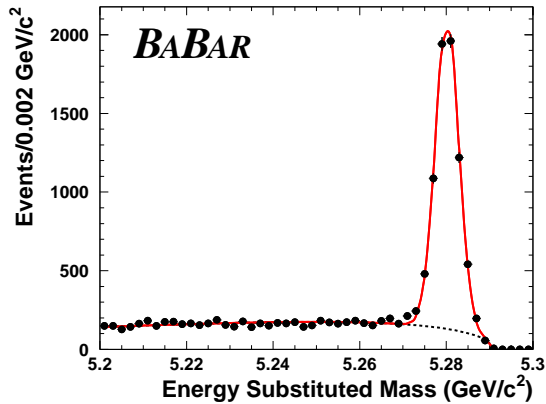}
\caption{Fit of background (dashed) and signal (solid) contributions to the
``energy-substituted mass'' (BaBar).}
\label{mes}
\end{figure}

The signal region is defined by constraints on $\Delta E = E^*_B -
E^*_{\rm beam}$ and  $M_{\rm bc} = \sqrt{E_{\rm
beam}^{*\,2}-{p_B^*}^2}$, where $E^*_B$ and $p_B^*$ are the
center-of-mass energy and momentum of the fully reconstructed $B$
candidate, and $E_{\rm beam}^*$ is the center-of-mass energy of the
beam. The background contributions can be fitted from $M_{\rm bc}$,
the ``energy-substituted mass'', as shown on Fig.~\ref{mes}.

\begin{figure}[htb]
\includegraphics*[width=0.8\hsize]{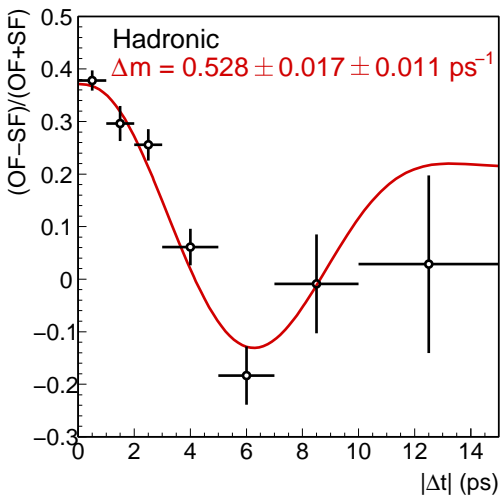}
\caption{Mixing asymmetry and result of the fit to hadronic modes (Belle).}
\label{had_fit}
\end{figure}

BaBar performs an unbinned likelihood fit including resolution
parameters on 30~fb${}^{-1}$ on-resonance data and finds
$\dmd = 0.516\pm0.016\pm0.010$. Belle uses the resolution function
used for lifetime measurements~\cite{lifetime} and obtains
$\dmd = 0.528\pm0.017\pm0.011$~ps${}^{-1}$ from 29~fb${}^{-1}$
on-resonance data (see Fig.~\ref{had_fit}).

\section{Summary and prospects}

\begin{figure}[htb]
\includegraphics*[width=\hsize]{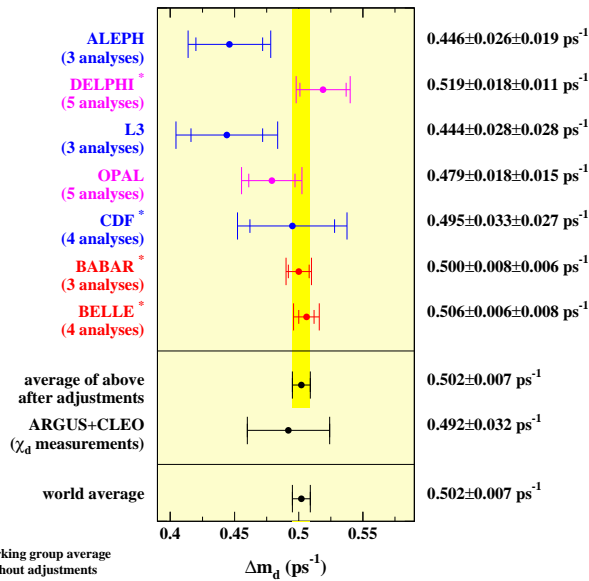}
\caption{Summary of $\dmd$ measurements and current world average.}
\label{PDG2003}
\end{figure}

Fig.~\ref{PDG2003} shows the most up-to-date summary of the results,
as selected for the 2003 issue of the PDG review. The world average is:
$\dmd = 0.502\pm0.007$~ps${}^{-1}$~\cite{PDG2004} (including statistical
and systematical errors), with an accuracy of $1.4\%$.

Within 3 years, BaBar and Belle will have collected 500~fb${}^{-1}$
each (about 15~times more than what was used here). An extrapolation of
the systematic errors for each individual measurement has been made in
order to estimate the accuracy on $\dmd$ that could be
reached. Improvement of the current limiting systematic errors are
evaluated from: a better precision on the $B$ lifetime, a larger amount
of Monte-Carlo statistics, an accurate measurement of some branching
fractions (e.g. $B\rightarrow D^{**}\,\ell\,\nu$). In addition,
resolution parameters are expected to be better extracted from
data. The average value is then computed in the same way as for the
world average, with the central value fixed to the present average. A
total error of $0.0023$~ps${}^{-1}$ is found, which corresponds to
$0.5\%$ of $0.502$~ps${}^{-1}$.

These extrapolations do not take into account possible improvements
of the existing analyses. On the other hand, the evolution of systematic
errors is hard to predict. This $0.5\%$ accuracy should therefore be
treated carefully. 

\section{Conclusion}

A number of measurements of $\dmd$ have been performed by the BaBar
and Belle collaborations. These efforts have lead to a world average
of $\dmd = 0.502\pm0.007$~ps${}^{-1}$. The error takes into account
statistical and systematical correlations between the measurements. The
current accuracy on $\dmd$ is currently 1.4\%, and is expected to
reduce to about half a percent within a few years.

In the future, $\Delta m_s$ will be measured with high precision.
The error on $\dmd$ may then become a limiting factor on the
determination of related CKM matrix parameters (see discussion 
in~\cite{YellowBook}, chapters~4 and~5).
In the meanwhile, as the accuracy on $\dmd$ is approaching the percent level, 
efforts are moving to more fundamental tests of underlying assumptions.
Limits on $CPT$ violating parameters have been set~\cite{bel_dilep},
and measurements of $\Delta\Gamma$ have started~\cite{DG}.


\end{document}